\documentclass[prd,preprint,a4paper,superscriptaddress,nofootinbib,11pt]{revtex4}

\usepackage{amsfonts}
\usepackage{amssymb}
\usepackage{hyperref}

\def\fs{\footnotesize}
\begin{document}

\title{\Large Accelerated Cosmological Models  in \\ First-Order Non-Linear Gravity}

\date{\today}

\author{Gianluca ALLEMANDI}
\email{allemandi@dm.unito.it}
\affiliation{{\fs Dipartimento di Matematica}, 
{\fs Universit\`a di Torino}\\{\fs Via C. Alberto 10, 10123
TORINO (Italy)}}
\author{Andrzej BOROWIEC}
\email{borow@ift.uni.wroc.pl}
\affiliation{\fs Institute of Theoretical Physics, University of Wroc{\l}aw\\
Pl. Maksa Borna 9, 50-204  WROC{\L}AW (Poland).}
\author{Mauro FRANCAVIGLIA}
\email{francaviglia@dm.unito.it}
\affiliation{{\fs Dipartimento di Matematica}, 
{\fs Universit\`a di Torino}\\{\fs Via C. Alberto 10, 10123
TORINO (Italy)}}


\begin{abstract}
The evidence of the acceleration of universe at present time has lead to
investigate modified theories of gravity and alternative theories of gravity,
which are able to explain acceleration from a theoretical viewpoint without
the need of introducing dark energy. 
In this paper we study alternative gravitational theories defined by  Lagrangians
which depend on general functions of the Ricci scalar invariant
 in minimal interaction with matter, in view of their
possible cosmological applications. Structural equations for the spacetimes
described by such theories are solved and the corresponding field equations
are investigated in the Palatini formalism, which prevents instability problems.
Particular examples of these theories are also shown to provide, under suitable
hypotheses, a coherent theoretical explanation of earlier results concerning
the present acceleration of the universe and cosmological inflation. 
We suggest
moreover a new possible Lagrangian, depending on the inverse of $\sinh(R)$,
which gives an explanation to the present acceleration of the universe.

\end{abstract}

\maketitle

\section{Introduction}
Recent astronomical observations have provided strong evidence that we live in an accelerating universe. The supernova  observation results (see for example  \cite{Perlmu} and \cite{Riess}), the observations about the anisotropy spectrum of the cosmic microwave background (CMBR) (see for example \cite{Spergel}) and the results about the power spectrum of large-scale structure \cite{Verde} have converted cosmologists to the acceptance of the evidence of an accelerating universe.\\
By itself, acceleration is easy to understand in the context of General Relativity as well as in quantum field theory; however problems and doubts on the correct theoretical model to
interpret observational data arise, owing to the very small but nonzero energy  scale which is seemingly implied. As a matter of facts, if we believe that we live in a universe which is homogeneous, isotropic and accelerating, General Relativity is unambiguous about the need for some sort of  \textit{dark energy} source to explain acceleration \cite{Carrol1}.
We are thus faced with some problems concerning: (i) the small amount of energy of the vacuum, which is much smaller than we estimate it to be (the so-called \textit{cosmological constant problem}); (ii) the nature of the dark energy which seems to dominate the universe and; (iii)  the \textit{coincidence problem} between the actual density of dark energy in the universe and the actual matter density \cite{Carrol1}. \\
The real nature of dark energy, which is required by General Relativity in this cosmological context,  is unknown but it is fairly well accepted that dark energy   should behave like a fluid with a large negative pressure. This could be explained in vacuum by means of a very small cosmological constant, which is in fact related with the energy of the vacuum, or by assuming the presence of some matter field, the so-called \textit{dark matter} or \textit{dark energy}. The dark energy models with effective equation of state $w_{eff}$ (which determines the relation between pressure $p$ and density of matter $\rho$) smaller than $w_{eff}<-1$ are currently preferrable, owing to the experimental results of  \cite{Spergel}. Other possibilities include a dynamical scalar field called \textit{quintessence} \cite{6meng}, or a phantom scalar field \cite{4meng, 5meng}, exotic perfect fluids \cite{7meng}, tachyon matter \cite{8meng}, a four Fermion model \cite{9meng} and a Born-Infeld Quantum condensate model \cite{10meng}.\\
One of the first attempts to correctly interpret, from a theoretical point of view, the observed data modifying General Relativity (without the introduction of dark energy) was to address the cosmological constant problem to somehow allow for the vacuum energy to be large \cite{Carrol3}; but this has been proven
to be not enough to get rid of dark matter and dark energy. \\
The other possibility is to assume that we do not yet understand gravity at large scales, which means that General Relativity should be modified or replaced by alternative gravitational theories when the curvature of spacetime is small (see for example  \cite{brane}). \\
\noindent Rather than solving the cosmological constant problem or introducing dark matter, we can try to explain the current period of acceleration of the universe by a modification of General Relativity which leads to modified Friedmann equations (MFR) so that the acceleration kicks in. MFR equations of this type arise surely in brane-world models with large spatial extra dimensions \cite{brane}.\\
In a completely different framework it has been realized  that in the quantization on curved spacetimes, when interactions among the quantum fields and the background geometry or the self interaction of the gravitational field are considered, the standard Hilbert-Einstein Lagrangian has to be suitably modified \cite{staro}. These corrective terms, which are essential in order to remove divergences, are higher-order terms in the curvature invariants such as $R$, $R^{\mu \nu} R_{\mu \nu}$, $R^{\mu \nu \alpha \beta} R_{\mu \nu \alpha \beta}$, $R\square^lR $, or non minimally coupled terms between scalar fields and the gravitational field. This is a further reason to support the idea that Lagrangians depending on general functions of the curvature invariants can provide physically significant models to explain cosmological experimental results. It is moreover interesting that such corrective terms to the standard Hilbert-Einstein Lagrangian can be predicted by some time-dependent compactification in string/M-theory (see \cite{capozziello}, \cite{Nojiri}). In particular it has been shown that quantum fluctuations in nearly flat spacetimes may induce terms which are   proportional to  inverse powers of the Ricci scalar invariant for small R, while the expansion of the effective action at large curvature predicts terms with a positive power of the curvature invariants \cite{Nojiri}, \cite{Nojiri1}.\\
As an alternative to extra dimensions, it is also possible to explain the modified Friedmann equations by means of a modified theory of four dimensional gravity \cite{capozziello}, \cite{carrol2}. A simple task to modify General Relativity, when the curvature is very small, is hence to add to the Lagrangian of the theory a piece which is proportional to the inverse of the scalar curvature $1 \over R$ (see  \cite{capozziello}, \cite{carrol2}, \cite{Nojiri}, \cite{Nojiri1},  \cite{Barraco}, \cite{Meng1},  \cite{Vollick}, \cite{Flanagan1}, \cite{Dick}   and references therein). These theories have been deeply analyzed in recent months, in view of their capability of explaining present cosmological  acceleration as well as early time inflation. As a matter of facts it appears that alternative theories of gravity provide values of $w_{eff}<-1$ and in accordance with the experimental results; see \cite{carrol2} and \cite{Nojiri1}.\\
We remark moreover that modified theories of gravity, depending on any analytical function of the Ricci scalar, have been studied in order to avoid the singularities in cosmological solutions \cite{kerner}. It is worth noticing that black holes in modified gravity are less entropic than in  standard General Relativity \cite{BNOV}.\\
 \\
It turns out that the metric approach to these Lagrangians leads to modified Friedmann equations which could explain the observed cosmological acceleration of the universe without the need of dark energy \cite{carrol2}. However the metric approach leads to complicated fourth order equations that can only be simplified by introducing fictitious scalar fields \cite{Magnano}. It has moreover been proven that the aforementioned   approach leads to results which are in contrast with the solar system experiments \cite{Chiba} and also that the relevant fourth order field equations suffer serious instability problems \cite{Dolgov}.\\
It has been shown in \cite{Vollick}  that, on the contrary, a Palatini variational approach to such modified nonlinear gravitational Lagrangians coupled to matter  produces second order field equations. These equations are equivalent to the standard Einstein field equations only in the vacuum case \cite{FFV} or for special  (radiating) matter. The modified Friedmann equations obtained in this context  offer however an alternative explanation for the cosmological acceleration and the approach to the de-Sitter  spacetime is exponentially fast when the $1 \over R$ term dominates. These modified Friedmann equations are not afflicted by instability problems and they are in acceptable accordance with the results of the solar system experiments \cite{Meng1}. The Palatini formalism has been further proven to be not excluded by electron-electtron scattering experiments provided the physical fields are taken into account in a suitable way  \cite{Magnano}, \cite{Vollick2}.\\
On the other hand it is widely believed and accepted that at the very early times the universe also underwent an acceleration phase called \textit{inflation}, the origin of which is still unknown.
Some authors have proposed that modified gravitational Lagrangians can moreover explain early inflation;  \cite{staro} and \cite{Nojiri}. The addition of correction terms of the
pure-power type $R^m$ ($m>0$) in addition to the $1 \over R$ term to the standard Hilbert-Einstein Lagrangian, can explain both the early time inflation and the current acceleration  without instability problems of field equations \cite{Nojiri1}. It is there proven that the terms in the Lagrangian with positive powers of the curvature support the inflationary period, while terms with negative powers of $R$ are able to explain the current cosmic acceleration. \\
\\
In this paper we first extend the results previously obtained in \cite{FFV} (for Lagrangians depending on general functions of the Ricci scalar invariant in  vacuum), to the more general case of interaction with matter fields. We consider a first order  Palatini formalism for field equations of a generalized gravitational Lagrangian in minimal interaction with a matter Lagrangian. The gravitational Lagrangian depends on a metric $g$ and a torsionless connection $\Gamma$, assumed to be a priori independent. The method generates structural equations for a spacetime with an ensuing conformal bi-metric structure.\\
 Starting from  the request that the connection should be metric with respect to a metric $h$ (suitably defined by dynamics) we are able to obtain \textit{generalized Einstein field equations}. The metric $h$ turns out to be related to the metric $g$ by a conformal transformation. These general results  extend those of  the vacuum case  \cite{FFV}; the universality of the Einstein field equations does not hold anylonger, apart from the case of a purely radiation universe (in this case the stress-energy tensor is traceless and the structural equations for spacetime are the same as in the vacuum case). The universality property was investigated not only at the level of  equations of motion, but also at the level of the definition of energy and conservation laws for the gravitational field. This problem was in fact analyzed  in the paper \cite{BFFV1}, where it was shown that universality holds also for the gravitational energy-momentum complex, which turns out to be essentially the same as in the linear case.  The formalism, developed in this paper  to obtain the generalized Einstein field equations, is very general and in forthcoming papers (under completion) it  will be moreover generalized to  theories depending on  higher-order terms in the Ricci-squared curvature invariant \cite{Ricci2}, which have been already analyzed in the  vacuum case in \cite {BFFV}.\\
For cosmological applications, we shall substitute  the general Robertson-Walker metric $g$ in the generalized Einstein field equations, assuming the stress-energy tensor to be a perfect fluid tensor. Taking into account the conformal relation between the (physical) metric $g$ and the conformal metric $h$, we obtain generalized Einstein equations which reproduce a \textit{modified Hubble constant}, depending on the conformal factor. We stress that, owing to the conformal transformation, also the metric $h$ results to be Robertson-Walker. The formalism is fully and analytically applicable under the request that the structural equation can be solved so that both $f(R)$ and $f'(R)$ can be expressed in terms of the trace  $\tau=T_{\mu \nu} g^{\mu \nu}$ (see also \cite{Meng1}). Otherwise, numerical and approximation methods are always applicable. These alternative theories are very rich in their structure and they can provide models of multi-universe cosmologies and can be suitably modified to fit experimental data.\\
Examples can be provided by choosing particular expressions of $f(R)$. We treat here the simplest cases of $f(R)=\beta R^n$ (pure power)  and  $f(R)=\alpha \; ln (\beta R)$ (logarithmic Lagrangian), which are very significant in view of their cosmological applications. The modified Hubble constant and the first order field equations are obtained, both for $h$ and $g$. We propose moreover a new possible Lagrangian, which deserves interesting cosmological applications, containing terms in the inverse of the hyperbolic sinus of the Ricci scalar.\\
The example $f(R)=\beta R^n$ is remarkably important. As a matter of facts each analytical Lagrangian in $R$ can be locally approximated by means of a Taylor polynomial expansion and further on polynomial Lagrangians in $R$ can be (under suitably hypotheses) locally approximated by $f(R)\simeq \beta R^n$, where n will be negative for small $R$ and positive for large $R$. 
In the aforementioned case of $f(R)=\beta R^n$ we specify the stress-energy tensor for the cases of dust, radiation and vacuum universes. In the dust universe and for the particular choice of zero space curvature in the Robertson-Walker metric we can solve exactly the differential equations  and we obtain the value of the acceleration parameter depending on the exponent $n$ (for particular values of $n$). It results that acceleration is supported in the cases $n<0$ and $n> {3 \over 2}$. These models seem to be able to explain both the current universe acceleration and early time inflation, taking into account leading terms of polynomial-like Lagrangians in each epoch of the universe. Polynomial Lagrangians are also discussed in order to treat radiation universes. It results that present time acceleration can be predicted and a critical value of the scale factor appears, which rules the transition from an accelerating to a decelerating epoch of the universe.\\
Logarithmic Lagrangians in the  Ricci scalar give an interesting, exactly solvable, model also for the case of pure radiation matter, supporting the present time acceleration of universe (see also \cite{Meng2} and \cite{NojiriO}).\\
Lagrangians containing terms proportional to the  inverse of the hyperbolic sinus of the Ricci scalar provide very complicated modified Friedmann equation which cannot be solved analytically. However these equations could give an explanation for the current acceleration of the universe in the appropriate limit and they can be useful for further numerical analysis or finer approximation techniques.\\


\section{$f(R)$ gravity}

\noindent We begin with considering on a $4$-dimensional Lorentzian manifold ($M,g$) the action
\begin{equation}
A=A^f_{\mathrm{grav}}+A_{\mathrm{mat}}=\int (\sqrt{g}f(R)+2\kappa L_{\mathrm{mat}})d^{4}x
\end{equation}
where $R\equiv R( g,\Gamma) =g^{\alpha\beta}R_{\alpha \beta}(\Gamma )$ and $
R_{\mu \nu }(\Gamma )$ is the Ricci tensor of any independent torsionless connection $\Gamma$.
The gravitational part of the Lagrangian is controlled
by a given real analytic function of one real variable $f(R)$,
while $\sqrt g$ denotes the scalar density
$\mid\det\parallel g_{\mu\nu}\parallel\mid^{\frac{1}{2}}$ of weight $1$.
The total Lagrangian contains also a first order matter part
$L_{\mathrm{mat}}$ functionally depending on matter fields $\Psi$ together with their first derivatives,
equipped with a gravitational coupling constant $\kappa=8\pi G$.\\
Equations of motion ensuing from the first order \'a la Palatini formalism
are  (see \cite{Barraco, Vollick,FFV})
\begin{eqnarray}
f^{\prime }(R)R_{(\mu\nu)}(\Gamma)-\frac{1}{2}f(R)g_{\mu \nu
}&=&\kappa T_{\mu \nu }  \label{ffv1}\\
\nabla _{\alpha }^{\Gamma }(\sqrt{g}f^{\prime}(R)g^{\mu \nu })&=&0
\label{ffv2}
\end{eqnarray}
where $T^{\mu\nu}=-\frac{2}{\sqrt g}\frac{\delta L_{\mathrm{mat}}}{\delta g_{\mu\nu}}$
denotes the matter source stress-energy tensor and $\nabla^{\Gamma}$ means covariant derivative with respect to $\Gamma$. In this paper  the metric $g$ and its inverse
are used for lowering and raising indices.\\
We shall use the standard notation
denoting by $R_{(\mu\nu)}$ the symmetric part of $R_{\mu\nu}$, i.e.
$R_{(\mu\nu)}\equiv \frac{1}{2}(R_{\mu\nu}+R_{\nu\mu})$. 
In order to get (\ref{ffv2}) one has to additionally assume
that $L_{\mathrm{mat}}$ is functionally independent of $\Gamma$; however it may
contain  metric  covariant derivatives $\nabla^g$ of fields.
This entails that the matter stress-energy tensor $T_{\mu\nu}=T_{\mu\nu}(g,\Psi, \nabla^g \Psi)$
depends on the metric $g$, the matter fields denoted
here by $\Psi$, together with their metric covariant derivatives. Physically speaking we are assuming
that matter fields are minimally coupled to the gravitational field $g$.\\
From (\ref{ffv2}) one sees that $\sqrt{g}f^{\prime }(R)g^{\mu \nu }$
is a symmetric twice contravariant tensor density of weight $1$, so that if
 it is not degenerate one can use it to define a metric $h_{\mu \nu}$ such that
the following holds true
\begin{equation}\label{h_met}
\sqrt{g}f^{\prime }(R)g^{\mu \nu}=\sqrt{h}h^{\mu \nu }
\end{equation}
This means that the two metrics $h$ and $g$ are conformally equivalent. The
corresponding conformal factor can be easily found (up to a possible sign) to be $f^{\prime}(R)$
(in $\dim M=4)$ and the conformal transformation results to be:
\begin{equation}\label{h_met2}
h_{\mu \nu }=f^{\prime}(R)g_{\mu \nu }
\end{equation}
Therefore, as it is well known, equation (\ref{ffv2}) implies that  $\Gamma =\Gamma _{LC}(h)$, i.e. the Levi-Civita connection of $h$, and $R_{(\mu\nu)}(\Gamma)=R_{\mu \nu }(h)\equiv R_{\mu\nu}$. We should emphasize, however, that scalar $R=g^{\mu \nu }R_{\mu \nu }(h)$
is not a Ricci scalar of the metric $h$. In what follows we shall call it a {\it generalized Ricci scalar} and sometime for shortcut simply a Ricci scalar.\\
Equation (\ref{ffv1}) can be supplemented by the scalar-valued equation
obtained by taking the $g$-trace of (\ref{ffv1}), where we set $\tau=\mathrm{tr} T=g^{\mu \nu }T_{\mu \nu }$:
\begin{equation} \label{struct}
f^{\prime }(R)R-2f(R)= \kappa\tau
\end{equation}
We obtain that equation (\ref{struct}) controls solutions of equation (\ref{ffv1}). We shall refer to this scalar-valued equation as the \textit{structural equation} of the spacetime.
More precisely, for any real solution $R=F(\tau)$ of (\ref{struct}) we have that  $f(R)=f(F(\tau))$ and $f^\prime(R)=f^\prime(F(\tau))$ can be seen as functions of $\tau$. For notational convenience we shall use the abuse of notation $f(\tau)=f(F(\tau))$ and $f^\prime(\tau)=f^\prime(F(\tau))$.\\
Now we are in position to introduce the generalized Einstein  equation under the  form
\begin{equation}\label{gen_Ein}
R_{\mu \nu }\left( h\right)=\frac{f(\tau)}{2f^{\prime}(\tau)}
 g_{\mu \nu }+\frac{\kappa }{f^{\prime}(\tau)} T_{\mu \nu }
\end{equation}
with $h_{\mu\nu}$ defined by (\ref{h_met2})
for a given $g_{\mu \nu }$ and $T_{\mu\nu}$
(see also \cite{Barraco,Vollick,FFV}). For the matter-free case we  find
that $R=F(0)$ becomes a constant, which implies that the two metrics are homotetic;
this property further implies that  equation (\ref{gen_Ein}) is nothing but Einstein
equation for the metric $g$ and it is almost independent on the choice of the function $f(R)$. This is nothing but the \textit{universality property} observed in \cite{FFV}.
Also Einstein equation with cosmological constant $\Lambda$ can be recasted into the form (\ref{gen_Ein}) if we chose  $f(R)=R-\Lambda$.
These properties justify the name of generalized Einstein equation for (\ref{gen_Ein}).
In the presence of matter (\ref{gen_Ein}) expresses instead a deviation for the metric $g$ to be Einstein,
apart from the case when $\tau=0$ where universality still holds true \cite{FFV}. \\
It should be also noticed that equation (\ref{struct})
has, in general, many real solutions, especially when $f(R)$ is a polynomial
function of higher degree. Therefore, $f(R)$-gravity supports the idea of multi-universe
interpretation.
Moreover, the number of possible universes is dynamical, since $\tau$ turns out to be a function of the FRW scale factor $a(t)$ (see below) except for a  radiation dominated period.\\
\noindent Another special property  one wants to emphasize is that we can make $f^{\prime }(R)R-2f(R)$ to be any function we wish. This is, of course, due to the fact that the non-homogenous linear ODE
\begin{equation}
f^{\prime }(R)R-2f(R)=\phi(R)
\end{equation}
has a solution in the form $f(R)=R^2\int ds\frac{\phi(s)}{s^3}$.
The choice of $f(R)$ allows to \textit{design} a specific cosmological scenario and to adjust the model to fit
concrete experimental data.


\section{FRW cosmology in $f(R)$ gravity}
\noindent For the cosmological applications one has to choose the metric $g$ to be the Friedmann-Robertson-Walker metric, which (in spherical coordinates) takes the standard form:
\begin{equation}
g=-d t^2+a^2 (t) \Big[ {1 \over {1-K r^2}} d r^2+ r^2 \Big( d \theta^2 +\sin^2 (\theta) d \varphi^2  \Big) \Big] \label{RW1}
\end{equation}
where $a (t)$ is the so-called \textit{scale factor} and $K$ is the space curvature ($K=0,1,-1$).
Another main ingredient of the cosmological model is the perfect fluid stress-energy tensor
$$T_{\mu\nu}=
(\rho+p)u_\mu u_\nu+pg_{\mu\nu}$$
where $p$ is the pressure, $\rho $ is the density of matter and $u^\mu$ is a co-moving fluid vector, which in a co-moving frame
($u^\mu=(1,0,0,0)$) becomes simply:
\begin{equation}
T_{\mu \nu}=
\left(
\begin{array}{clcr}
\rho&0&0&0\\
0&\frac{pa^2 (t)}{1-K r^2}&0&0\\
0&0&pa^2 (t)r^2&0\\
0&0&0&pa^2 (t)r^2\sin^2 (\theta)
\end{array}
\right) \label{Tmunu}
\end{equation}
Later on we shall assume the standard relations between the pressure $p$, the matter density $\rho$
and the expansion factor $a(t)$, namely:
\begin{equation}
p=w\rho \quad , \quad
\rho=\eta a^{-3(1+w)} \label{pro}
\end{equation}
where particular values of the parameter $w\in \{-1,0,{1\over 3}\}$ will correspond to the
vacuum,  dust or radiation   dominated universe. Exotic matters, which are up to now under investigation as possible models for dark energy, admit instead values of $w<-1$. These  expressions (\ref{pro}) follow  from the conservation law of the energy-momentum  $\nabla^\mu T_{\mu \nu}=0$ and consequently  the continuity equation  should hold:
\begin{equation}
\dot{\rho}+3H(\rho+p)=0
\end{equation}
where $H=\frac{\dot a}{a}$ is the \textit{Hubble constant}.
The requirements  (\ref{ffv2}) and (\ref{h_met})  fix $h$ to be  conformal to $g$
and  in fact equal to:
\begin{equation}
h= f^\prime(\tau)\Big{\{}-d t^2+a^2 (t) \Big[ {1 \over {1-K r^2}} d r^2+ r^2 \Big( d \theta^2 +\sin^2 (\theta) d \varphi^2  \Big) \Big] \Big{\}} \label{VolR1}
\end{equation}
where  $\tau=T_{\mu \nu} g^{\mu \nu}=3p-\rho$ is  a function of time through its dependence on the scale factor $a(t)$: $$\tau=(3w-1)\eta [a(t)]^{-3(1+w)} \ .$$
Substituting all necessary ingredients into the generalized Einstein equation (\ref{gen_Ein}) we obtain the following
\begin{equation}
R_{00}(h)=\frac{-f(\tau)+2\kappa\rho}{2b}=
-\frac{3}{2}\Big[2\frac{\ddot a}{a}+\frac{\dot b}{b}\,\frac{\dot a}{a}
+\frac{\ddot b}{b}-\Big(\frac{\dot b}{b}\Big)^2 \Big]\label{00comp}
\end{equation}
for the $00$ component, while for the  $11$ component we have:
\begin{equation}
R_{11}(h) \left[ \frac{1-K r^2}{a^2} \right]=\frac{f(\tau)+2\kappa\,p}{2b}=
\frac{1}{2}\Big[2\frac{\ddot a}{a}+5 \frac{\dot b}{b}\,\frac{\dot a}{a}+\frac{\ddot b}{b}+4\Big(\frac{\dot a}{a}\Big)^2 +4\frac{K}{a^2}\Big]\label{11comp}
\end{equation}
Here for simplicity we have introduced the shortcut $b(t)=f^\prime(\tau)$. Of course for the usual
 Einstenian case one has $b(t)=1$. Combining the last two equations we can obtain
 an analogue of the Friedmann equation under the form
\begin{equation}
\Big(\frac{\dot a}{a} +\frac{\dot b}{2b}\Big)^2+ \frac{K}{a^2}=
\frac{\kappa}{3b}\Big[\rho+\frac{f(\tau)+\kappa\tau}{2\kappa}\Big] \label{MHC}
\end{equation}
which can be seen as a generalized definition of a \textit{modified Hubble constant} $\hat{H}=\Big(\frac{\dot a}{a} +\frac{\dot b}{2b}\Big)$,
which takes into account the presence of the conformal factor $b$, entering the
definition of the conformal metric $h$, which in turn generalizes the results of \cite{Meng1}.\\
We stress that, owing to the conformal transformation between $h$ and $g$, the generalized Ricci scalar can be generally expressed from (\ref{00comp}) and (\ref{11comp}) as follows:
\begin{equation}
R=R_{\mu \nu} (h) g^{\mu \nu}=\frac{2 f(\tau)+ \kappa \tau}{b}=R(h) b
\end{equation}
which reproduces General Relativity in the particular case $f(R)=R$, as we shall see later on, but provides relevant deviation from General Relativity in the other cases. We remark that R(h) is the true Ricci scalar of the metric $h$.
\noindent As it has been already remarked in \cite{Barraco} the metric $h$ itself is a Robertson-Walker
metric (see (\ref{VolR1}),
\begin{equation}
h=\epsilon \left( -d \tilde{t}^2+A^2 (\tilde t) \Big[ {1 \over {1-K r^2}} d r^2+ r^2 \Big( d \theta^2 +\sin^2 (\theta) d \varphi^2  \Big) \Big] \right) \label{RW2}
\end{equation}
with a new cosmic time $d\tilde t=\sqrt{|b|}dt$ and a new scale factor $A=\sqrt{|b|}a$,
where $\epsilon=\pm 1$ corresponds to positive or negative values of $b$.
The generalized Einstein equation (\ref{gen_Ein}) can be also calculated in
$({\tilde t}, x^i)$- coordinates. This is equivalent to  the assumption
that the metric $h$ is a physical one (i.e., that we can use conformal frame instead of the original Einstein frame).  We  obtain in this case:
\begin{equation}
\epsilon\frac{-f(\tau)+2\kappa\rho}{2b^2}=
-3\frac{\ddot A}{A}\label{00comp2}
\end{equation}
for the $00$ component while for the $11$ component we find
\begin{equation}
\epsilon\frac{f(\tau)+2\kappa\,p}{2b^2}=
\frac{\ddot A}{A}+2\Big(\frac{\dot A}{A}\Big)^2 +2\frac{K}{A^2}\label{11comp2}
\end{equation}
where $\dot A$ denotes now the differentiation with respect to the new
cosmic time $\tilde{t}$.
An additional factor $\frac{\epsilon}{b}$ on the l.h.s. appears due to the fact
that $dt^2=\frac{1}{|b|}d\tilde{t}^2$ and $a^2=\frac{1}{|b|} A^2$.
Now the analogue of the Friedmann equation  takes the form
\begin{equation}
\tilde{H}^2 =- \frac{K}{A^2}+\epsilon
 \frac{f(\tau)+\kappa(\rho+3p)}{6b^2} \label{MHC2}
\end{equation}
with $\tilde{H}=\frac{\dot A}{A}$ being the Hubble constant of the conformal metric $h$. This expression is much
simpler and recalls the standard Friedmann equations for the Robertson-Walker metric.


\subsection{Examples: $f(R) =\beta R^n$ and polynomial  Lagrangians}

\noindent We consider, as a class of particular examples, the class of  linear Lagrangians  in  an arbitrary power of the curvature invariant $R$.
The importance of such models can be considered and understood in connection with the main effort of modifying General Relativity by means of
alternative Lagrangians, which are able to explain  the experimental data in some limit. Up to now (and up to our knowledge) polynomial-like and logarithmic-like Lagrangians
in the generalized Ricci scalar invariant have been considered (see for example \cite{Carrol1}, \cite{capozziello}, \cite{carrol2}, \cite{Nojiri},   \cite{Meng1}, \cite{Vollick},  \cite{Meng2}, \cite{NojiriO}  and references therein). These particular Lagrangians recover General Relativity in some approximation, which physically speaking means at some age of the universe, and reproduce modified Friedmann equations in some other limit. These modified Friedmann equations are able to give a possible  theoretical explaination to the  experimental results (such as inflation and present acceleration of the universe). \\
If we narrow down our researches to the case of polynomial
Lagrangians, each term of the Lagrangian behaves as a leading term
at some particular age of the universe and it is able to reproduce
at a convenient order of approximation the experimental data.
Moreover any analytical function can be approximated by means of
 its Taylor polynomial expansions in the limits of physical relevance and it consequently 
behaves like a polynomial-like Lagrangian.\\
In this framework it is hence worth analyzing exactly and
analytically Lagrangians of the type $L(R)=\beta R^n \sqrt g$
(with arbitrary possibly non-integer $n$) representing single
terms of a more physical and general polynomial or polynomial-like
expression for the Lagrangian. Integer values of $n$ assume a
fundamental role in this context representing, for any analytical
function, the terms deriving from a Taylor expansion. The
advantage which derives from this particular class of Lagrangians
is that they are easily and exactly solvable and they provide
coherent models for the
universe acceleration and for the early time inflation.\\
We stress once more that we are specializing to the case of four
dimensional spacetimes so that $g^{\mu \nu} g_{\mu \nu}=4$. We
assume then a  Lagrangian of the form:
\[
f(R)\sqrt{g} =\beta R^n \sqrt g \qquad (\beta \ne 0; \; n \in  {I\kern-.36em R}  ; \; n\ne 0,2)
\]
(see also \cite{Nojiri1} and \cite{Meng1}).
The algebraic field equations (\ref{ffv1}) are consequently the following:
\[
\beta R^{n-1} \Big[ n R_{\mu \nu}-{1 \over 2} R g_{\mu \nu} \Big]=\kappa T_{\mu \nu}
\]
while the structural equation (\ref{struct}) becomes:
\begin{equation}
R^n=\frac{\kappa \tau}{\beta(n-2)} \label{aa2}
\end{equation}
This expression implies that the case $n=2$ is  singular  (see e.g. \cite{Nojiri1}).
Taking into account the expression  (\ref{Tmunu}) for the stress energy tensor and  the dominant energy condition we should impose
\begin{equation}
\tau=T_{\mu \nu} g^{\mu \nu} <0
\end{equation}
Equation (\ref{aa2}) forces $R^n$ to be positive definite for even integer values of $n$ and  we should fix the dimensional
constant $\beta$ (the dimension of $\beta$ is the same as the dimension of $R^{1-n}$)
in front of the Lagrangian to be :
$$
\cases{
\beta>0 \quad \hbox{for} \quad n<2\cr
\beta<0 \quad \hbox{for}  \quad n>2
}
$$
Let us introduce, for convenience, a new re-scaled dimensional
coefficient $\tilde{\beta}=\beta(2-n)$ which is positive for even
$n$ and arbitrary otherwise. In the last case, i.e. for $n$ odd,
one has ${\rm sign} R={\rm sign} \tilde\beta$, where
$R=(-\kappa\tau/\tilde\beta)^{1\over n}$. Instead, for $n$ even
one has two different real solutions of (\ref{aa2}), namely:
\begin{equation}\label{aa3}
R_{\pm}=\pm \Big(\frac{-\kappa\tau}{\tilde\beta}\Big)^{1\over n}
\end{equation}
This implies that in any case  $R$ will be proportional to a well-defined power of $a$,
which is exactly: $R \simeq a^{-\frac{3(w+1)}{n}}$. This expression shows that only in the vacuum case the model approaches a de-Sitter (anti de-Sitter) universe.\\
The requirements (\ref{ffv2}) and (\ref{h_met}) fix $h$ to be  conformal to $g$
and in fact equal to:
\begin{equation}
h=\frac{n\varepsilon}{2-n}\tilde{\beta}^{1 \over n}( -\kappa \tau)^{\frac{n-1}{n}} \Big{\{}-d t^2+a^2 (t)
\Big[ {1 \over {1-K r^2}} d r^2+ r^2 \Big( d \theta^2 +\sin^2 (\theta) d \varphi^2  \Big) \Big] \Big{\}} \label{VolR2}
\end{equation}
where we remark again that $\varepsilon={\rm sign} R = 1$ for odd values of $n$ and, on
the contrary, $\varepsilon=\pm 1$ in accordance with the choice of the
solution in (\ref{aa3}) for even values of $n$. We are now able to calculate the modified Friedmann equations and the modified Hubble constant from (\ref{MHC});
or, in a different but completely equivalent way, by inserting (\ref{VolR2}) into (\ref{gen_Ein}). From the structural
equation (\ref{struct}) we can calculate:
$$
\cases{
f(\tau)=\frac{\kappa \tau}{n-2}\cr
b(t)=f^\prime(\tau)=\frac{n\varepsilon}{2-n}\tilde{\beta}^{\frac{1}{n}}( -\kappa \tau)^{\frac{n-1}{n}}
}
$$
and we obtain that the Hubble constant for the metric $g$ can be locally calculated to be:
\begin{eqnarray}
H^2= \left( \frac{\dot a}{a} \right)^2 = \frac{2n\varepsilon}{3(3w-1)[3w(n-1)+(n-3)]}
\left[  \frac{-\kappa \tau}{\tilde\beta} \right]^{\frac{1}{n}} 
-\frac{K}{a^2} \left[ \frac{2n}{3w(n-1)+(n-3)} \right]^2  \label{MFRn}
\end{eqnarray}
The deceleration parameter can be obtained from the Hubble
constant by means of the following relation:
\begin{equation}
q(t):=- \left(1+\frac{\dot{H}(t)}{H^2(t)} \right)=-\left(
\frac{\ddot{a}(t)}{a(t) H^2(t)} \right) \label{qhub}
\end{equation}
and from (\ref{MFRn}) it results, in the case $K=0$, to be formally equal to:
\begin{equation}
q(n, w)=\frac{3(1+w)-2n}{2 n} \label{qquti}
\end{equation}
We say formula (\ref{qquti}) to be \textit{formal} since, at the moment,
we do not know about its effective solution realizations. This will be the subject of our
investigations below.\\
The effective $w_{eff}$ can be obtained (as in \cite{carrol2}) by means of simple calculations from
 (\ref{qquti}) and it results to be, for this theory:
\begin{equation}
 w_{eff}={2\over 3} q(n, w)-{1 \over 3}=-1+{1 \over n}+{w \over n}\label{weff}
\end{equation} 
We remark that the range of $-1.45<w_{eff}<-0.74$ for dark matter, stated in \cite{Spergel}, can be
 easily recovered in our theory by choosing suitable and admissible values of $n$.\\
If we consider expression (\ref{qquti}) for the deceleration parameter $q(n, w)$ we see that, from the 
definition of $w_{eff}$ in (\ref{weff}), an accelerating behaviour of the cosmological model requires that the effective $w_{eff}<-{1 \over 3}$. This bound to the value of $w_{eff}$ can be alternatively seen in terms of $w$. This states in particular that to obtain accelerated universes we should else  impose $w <{2n \over 3 }-1$ for $n>0$ or $w >{2n \over 3 }-1$ for $n<0$. \\
For the case $n=1$ this imposes an upper bound for the acceleration to the value of $w<w_{crit}=-1$, reproducing the well-known results of General Relativity; choosing alternative Lagrangians with $n>0$ implies that this limit is shifted to the value $w_{crit}={2n \over 3 }-1$. In the case $n<0$ we have  no longer an upper bound, but a lower bound for  $w>w_{crit}={2n \over 3 }-1$; dust and radiation matter  satisfy that condition by definition.\\

\noindent Consider now, in more detail,  the case of pressure-free (dust) universe, i.e.
$p=0$, so that $\tau=-\rho(t)=-\frac{\eta }{ a^3 (t)}$
(we remark once more that $\eta >0$). The resulting generalized
Einstein equations (\ref{MFRn}) which derive from (\ref{ffv1}) produce modified
Friedmann equations in the matter universe case  considered here.
The resulting expression for the Hubble constant of $g$ can be
calculated from (\ref{MFRn}) to be (for $n \ne 3$):
\begin{equation}
H^2=\frac{\dot{a}^2(t)}{{a}^2 (t)}= {2 \over 3} \left[ \frac{\varepsilon n
(\eta \kappa )^{1 \over n}  }{(3-n) [\tilde\beta a^3 ]^{1 \over n} }-
\frac{6 K n^2}{a^2 (t) (n-3)^2} \right]
\label{FRV1}
\end{equation}
which reproduces, as it should be expected, the standard Friedmann equation in the very particular
case $n=1$, and $\tilde\beta=\beta=1$. We remark that the above expression (\ref{FRV1}) represents
the square value of the Hubble constant and it becomes singular at $n=3$ in the case of dust universe, under analysis.\\
We restrict now ourselves to the case of K=0. We have to require
that the expression on the right hand side of (\ref{FRV1})  should
be positive. \\
We start discussing  integer values of $n$, which are very relevant for our analysis, as we stated before. In the  case of odd integer values of $n$
($\varepsilon=1$) we should require $\tilde\beta>0$ for $0<n<2$ or
$2<n<3$, i.e. essentially in the case $n=1$ of standard General Relativity.
When $n$ is still odd but $\tilde\beta<0$ one gets $n<0$ or $n>3$ instead.\\
In the case of even values of $n$, we see that $\tilde\beta$ is always
positive. Therefore, as we required before, we should fix
$\varepsilon=-1$ for the following values of $n$: $n<0$ or $n>3$. Thus the solution
$R_+$ of (\ref{aa3}) gives no relevant contribution to the solutions of (\ref{FRV1}), owing the singularity in the case $n=2$.
Finally, one should notice that $\beta>0$  for $n\in
(0, 2)\cup(3,+\infty)$  and negative otherwise, i.e. for $n\in
(-\infty, 0)\cup(2, 3)$ independently of the parity of $n$. \\
We remark however that we do not need to assume that $n$ is an
integer and, in fact, it can be a priori any real (or rational) number. In any case, however, we should require equation (\ref{aa2}) to be definite and morever we should impose the positivity 
of (\ref{MFRn}). Suitable values of $\beta$ have to be chosen and the analysis can be carryed over, following the headlines of the analysis previously done in the case of integer values of $n$. Specializing again to the case of dust universe with $K=0$, in the case of real, non-rational $n$ one has to assume $\tilde{\beta} >0$ and this imposes $n \in (0,3)$.

\noindent The  modified Friedmann equations, for dust universe in the case $K=0$,
can be integrated and we easily obtain (apart from integration
constants, which can be forgotten in our analysis):
\begin{equation}
a(t)=\left[\frac{3\varepsilon}{2n(3-n)}\right]^{{n}\over{3}}
\left[\frac{\kappa \eta }{\tilde\beta}\right]^{\frac{1}{3}}   
t^{{2n} \over 3} \label{aa1}
\end{equation}
Now the deceleration parameter can be obtained again from (\ref{qquti}); we find:
\begin{equation}
q(t)=\frac{3-2n}{2 n} \label{quti}
\end{equation}
for the particular solutions (\ref{aa1}) corresponding to  the case
$K=0$, selected by (\ref{FRV1}). 
This implies that we obtain accelerated solutions in the case:
\begin{equation}
q(t)<0 \Leftrightarrow n<0\; \;\; \mbox{or} \;\; n>{3 \over 2}
\end{equation}
corresponding to alternative theories of gravity with Lagrangians
depending on inverse powers of the generalized Ricci scalar or on terms with
powers higher than $3\over 2$, owing to the restriction imposed for the positivity
of (\ref{FRV1}). \vskip5pt
\noindent If we consider the case of pure-radiation universe, corresponding to $\tau=3p-\rho=0$
with $\rho=\rho(t)=\frac{\eta}{a^4(t)}$, we have that   the equation (\ref{gen_Ein}) is undefined, since  (\ref{aa2}) implies $R^n=0$ and $f^\prime(\tau)=0$. This entails that our formalism fails to cover this case which has to be treated differently (see below).
Therefore, formula (\ref{qquti}) for $w=\frac{1}{3}$ is purely formal.\\

\noindent In vacuum dominated universe, corresponding to the case $w=-1$, we obtain from
(\ref{MFRn}) that the Hubble constant is:
\begin{equation}
H^2=\frac{\dot{a}^2(t)}{{a}^2 (t)}= {\varepsilon \over 12} \left[ \frac{4 \kappa \eta }{\tilde\beta }
\right]^{\frac{1}{n}}-\frac{ K }{a^2 (t)}
\label{FRV2}
\end{equation}
which results to be positive and well-defined for any $n$ ($n \ne 2$) with a suitable choice for
$\beta$ and $\varepsilon$. It follows that in the case of a vacuum universe,  and for $K=0$ we have that:
\begin{equation}
a(t)=e^{\sqrt{\frac{\varepsilon}{12} \left[ \frac{4 \kappa \eta}{\tilde\beta}\right]^\frac{1}{n}} t}
\end{equation}
corresponding to a de Sitter universe (with a power-law expansion);
the deceleration parameter is in this case $q(t)=-1$ supporting cosmological acceleration.\\

\noindent
 A similar analysis can be performed under the hypothesis that the metric $h$ is the physical
metric and it is spatially flat ($K=0$). The deceleration parameter for the metric $h$ can be formally calculated to be:
\begin{equation}
\tilde{q}(w,n)= \frac{-f(\tau)+2\kappa\rho}{f(\tau)+\kappa(\rho+3p)}=\frac{2n-3(1+w)}{n-3+3(n-1)w}
=\frac{2n-3(1+w)}{n(1+3w)-3(1+w)}
\end{equation}
which gives acceleration $\tilde{q}(-1,n)=-1$ for any value of $n$ in vacuum dominated
universes and  $\tilde{q}(0,n)=\frac{2n-3}{n-3}<0$ in dust dominated universes 
provided that $\frac{3}{2}<n<3$ ($n\neq 2$). Finally, for pure radiation we have $\tilde{q}(\frac{1}{3},n)=1$, i.e. one gets deceleration for any value of $n$.\\
The effective value of $w$ can be also calculated in this case  from (\ref{weff}); we get:
\begin{equation}
w_{eff}=\frac{4n-9(1+w)+n(1+3w)}{3[n(1+3w)-3(1+w)]}
\end{equation}
which allows us to make a comparison with the experimental data.\\
These results are alternative to the case described by formula (\ref{qquti}) when $g$ is the physical metric. As usual in alternative non-linear theories, we do not know a priori which is the physical metric: discussions about the physical interpretation of $g$ and $h$ both from the mathematical and physical viewpoint are up to now open; see, e.g. \cite{Magnano}, \cite{Vollick2} and \cite{new}.


\subsubsection{Polynomial Lagrangians in the generalized Ricci scalar}

\noindent As we told before theories with power Lagrangians in the generalized Ricci scalar can
be considered as \textit{approximations}
of more physical polynomial-like Lagrangians \cite{Nojiri} of the type:
\begin{equation}
f(R)=R+\frac{\alpha}{(2+n)R^n}+\frac{\beta}{2-m} R^m
\end{equation}
(here both $n>0$ and $m>0$, with $m \ne 2$ and $n \ne -2$).\\
We consider the case of dust universe $w=0$. In the limit of small or large curvatures,
corresponding to the cases of present time universe and early time universe,
 we obtain from the structural equations  that the leading terms are respectively:
$$
\cases{
R \rightarrow 0 \Rightarrow \frac{-\alpha}{R^n}=\kappa \tau\cr
R \rightarrow \infty \Rightarrow  -\beta R^m=\kappa \tau}
$$
From (\ref{quti}) we deduce that polynomial Lagrangians provide an
explanation for early time inflation assuming that $m> {3\over 2}$
and they can provide an explanation to present time cosmic
acceleration assuming that  some inverse power of the generalized Ricci scalar
is present in the Lagrangian (i.e. $\alpha \ne 0$). \\
This result reproduces previous results which have been obtained in a different framework both in the metric formalism (in \cite{carrol2} and \cite{Nojiri1}) and in the Palatini formalism (\cite{Meng1} and \cite{Vollick}). We stress moreover that terms like  $\beta R^m$  (with $m>0$) are related to the so-called Starobinsky inflation \cite{staro}.


\subsubsection{Polynomial Lagrangians for radiation universes}

\noindent We can otherwise consider, for the case of radiation
universes $w={1 \over 3}$, a Lagrangian of the form:\footnote{$m\neq 1,2$}
\begin{equation}
f(R)=\alpha R+\frac{\beta }{m-2}R^{m}  \qquad (\hbox{where}  \; \;w=\frac{1}{3}; \; \tau
=0)
\end{equation}
which  in \cite{Nojiri1} and \cite{Meng1} has been already examined in the case $m=2,\tau \neq 0$. It has   been there proven that,   in the metric formalism such Lagrangians support inflation, while in the Palatini formalism they provide explanation for the present time acceleration \cite{Meng1}.\\
The structural equation (\ref{struct}) admits besides the trivial
solution $R=0$ also a non-trivial solution
$R=\varepsilon {\frac{\alpha}{\beta}}^\frac{1}{1-m}$, where $\varepsilon =1$ for $m$ odd and $\varepsilon =\pm 1$ for even values of  $m$. This correspond to de-Sitter and anti-de Sitter universe for positive and negative values of $R$ respectively. \\

\noindent In the case of the obvious solution $R=0$ we have that $f(R)=0, \; \; f^{\prime}(R)=\alpha $, and consequently:
\begin{equation}
H^{2}=\frac{\kappa \rho }{3\alpha } =\frac{\kappa \eta }{3\alpha } a^{-4} - K a^{-2}
\end{equation}
The deceleration parameter can be formally calculated to be:
\begin{equation}
q(t)=\frac{\frac{\kappa \eta }{3\alpha }}{\frac{\kappa \eta }{3\alpha }-K a^{-2}}
\end{equation}
In the particular case $K=0$ we have $a=\sqrt[4]{\frac{\kappa \eta }{3\alpha }}\sqrt{2t}$ and the deceleration parameter can be easly calculated to be $q(t)=1$, describing a decelerating universe for this solution ($R=0$).\\
\noindent If we shift to the non trivial solution $R=\varepsilon \left( \frac{\alpha }{\beta }\right) ^{\frac{1}{
m-1}}$ we now see   that  the structural equation  gives:
\begin{eqnarray}
f(R)&=&\varepsilon \frac{m-2+\alpha }{m-2}\left( \frac{\alpha }{\beta
}\right) ^{\frac{1}{m-1}}\\
f^{\prime}(R)&=&2\alpha \frac{m-1}{m-2}
\end{eqnarray}
so that the Hubble constant for the metric $g$ can be calculated to be:
\begin{equation}
H^{2}=\frac{\kappa \eta \left( m-2\right) }{6\alpha \left( m-1\right) }
a^{-4}+\varepsilon\frac{m-2+\alpha }{12\alpha \left( m-1\right) }\left( \frac{\alpha }{
\beta }\right) ^{\frac{1}{m-1}}-K a^{-2}=\Sigma-K a^{-2}+\Lambda a^{-4}
\end{equation}
where for notational convenience we have introduced the parameters $\Lambda=\frac{\kappa \eta \left( m-2\right) }{6\alpha \left( m-1\right) } $ and $\Sigma=\varepsilon\frac{m-2+\alpha }{12\alpha \left( m-1\right) }\left( \frac{\alpha }{
\beta }\right) ^{\frac{1}{m-1}}$, which we require to be positive.\\
The deceleration parameter can be obtained by means of  formula (\ref{qhub}):
\begin{equation}
q(t,m)=\frac{\Lambda a^{-4}-\Sigma }{\Sigma -K a^{-2}+\Lambda a^{-4}}=\frac{\Lambda
-\Sigma a^{4}}{\Lambda -K a^{2}+\Sigma a^{4}}
\end{equation}
In the limit of large density matter $\rho$, corresponding to early time universes, we obtain  $q=1$;  on the contrary at late times we have $q=-1$ when $\rho $ results to be very small and forgettable. This corresponds to a presently accelerating universe.
We remark that the critical value \ $q=0$ \ (corresponding to a change from a deceleration epoch  to an accelerated epoch) is met when the radius attains the critical value:
\[
a_{c}=\sqrt[4]{\frac{\Lambda }{\Sigma }}=\sqrt[4]{\frac{2\varepsilon \kappa
\eta \left( m-2\right) }{m-2+\alpha }\left( \frac{\beta }{\alpha }\right) ^{
\frac{1}{m-1}}}
\]
We remark that both asymptotic values of $q$ and the critical value $a_c$ do not depend on the value of
 the spatial curvature $K$ of the spacetime under consideration.\\
This simple example illustrates two important properties of polynomial Lagrangians, which are full in physical significance:
\begin{itemize}
\item[-] there may exist two (or more) parallel universes, corresponding to different solutions of the structural equations of the same Lagrangian and matter source. These  different solutions provide models for different cosmologies;
\item[-] there exist solutions which provide a smooth transition from deceleration epochs to accelerated universes. This happens in correpondence with some critical value of the cosmic radius $a_c (t)$ and it is in relation with the so-called \textit{cosmic speed-up} \cite{carrol2}. 
\end{itemize}


\subsection{Another example: $f(R)=\alpha \ln (R)$  Lagrangians}
\noindent We chose as a further particular example the case $f(R)=\alpha \ln (\beta R)$,
which is  relevant
since logarithmic terms in the Ricci scalar are induced by quantum effects in curved spacetimes;  see \cite{Nojiri} and  \cite{NojiriO}.
Notice that the corresponding dimensions should be $[\alpha]=\frac{1}{[\beta]}=[R]$. For notational convenience we will fix $\beta=1$ in units  such that  dimensions  remain correct.\\
We obtain from the structural equation (\ref{struct})  that:
$$
\cases{
f(\tau)=\frac{\alpha-\kappa \tau}{2}\cr
b(t)=f^\prime(\tau)=\alpha e^{\frac{\kappa \tau-\alpha}{2 \alpha}}
}
$$
Substituting this expressions into (\ref{MHC}) and performing straightforward calculations the explicit expression for the Hubble constant can be easly obtained:
\begin{eqnarray}
H^2=\frac{1}{\Big[ 1-\frac{3 \kappa (1+w) }{4 \alpha} \tau  \Big]^2 } \left[ \frac{3 (w+1) \kappa \tau+\alpha (3w-1)}{12  \alpha (3w-1 )e^{\frac{\kappa \tau-\alpha}{2 \alpha}}}-\frac{ K}{a^2}   \right]
\label{MHClnR}
\end{eqnarray}
where we recall that $\tau$ can be expressed in terms of $a(t)$ as:
\begin{eqnarray}
\tau=3p-\rho=\eta (3w-1) [a(t)]^{-3(1+w)}
\end{eqnarray}
We remark that the Hubble constant can be specialized to
the case of dust, radiation or vacuum universes by a suitable choice of $w$ ($w=-1, 0,{1 \over 3}$). It follows from formula (\ref{MHClnR}) that the pure radiation case is singular and cannot be treated in our formalism. \\
In the case of dust universe ($w=0$) we have that:
\begin{eqnarray}
H^2=\frac{1}{\left( 1+\frac{3 \kappa  }{4 \alpha} \rho  \right)^2} \left[\frac{\alpha+3 \kappa \rho}{12 \alpha  } e^{\frac{\alpha+\kappa \rho}{2 \alpha}}-\frac{ K}{a^2}\right]
\label{Hlndust}
\end{eqnarray}
We consider the limits of small and large energy density in the universe, which are respectively $\alpha >>\kappa \rho $ and $\alpha << \kappa \rho $. In the case of small energy density, corresponding to late times of the universe, we have that:
\begin{eqnarray}
H^2=\frac{\sqrt{e}}{12}
\end{eqnarray}
which reproduces a power-law expansion with a universe approaching a de-Sitter universe. On the other hand, in the case of large density matter we have that, apart from constant and positive\footnote{the positivity is necessary and it follows from the positivity of $H^2$.} factors, the following holds $H^2 \simeq e^\rho$ and  this surely supports decelerating cosmological models. \\
The deceleration parameter can be moreover calculated formally. It turns out  to be, from (\ref{qhub}):
\begin{eqnarray}
q(t,w)&=&-1+\frac{9 \kappa (1+w)^2  \tau}{3 \kappa (1+w)\tau -4 \alpha }+\\&-&{1 \over 2}
\left[ \frac{3 \kappa (w+1) \tau[3(w+1) \kappa \tau+\alpha (-3 w-7)]}{24 \alpha^2 (3w-1) e^{\frac{\kappa \tau-\alpha}{2 \alpha}}}+ \frac{2 K}{a^2}\right]  
 \left[ \frac{3(w+1) \kappa \tau+ \alpha (3 w-1)}{12 \alpha (3w-1) e^{\frac{\kappa \tau-\alpha}{2 \alpha}}}- \frac{K}{a^2} \right]^{-1} \nonumber
\end{eqnarray}
which  can be suitably approximated in the cases of small and large energy densities. \\
We obtain that in the limit of large energy density, corresponding to early time universes, we will have  $q(t) \simeq 3(w+1)-2$, which provides decelerating universes, apart for the case of vacuum space. The limit of late time universe, namely for small energy density $\tau$, provides as expected from (\ref{Hlndust}) a de-Sitter like universe with $q(t)=-1$, which always provides  an accelerating cosmological model.\\
\noindent The analysis performed above implies that, in the Palatini formalism, Lagrangians proportional to the logarithm of the Ricci scalar provide cosmological models without an inflationary epoch, while these models are able to explain the current acceleration of the universe; see also \cite{Meng2} and \cite{NojiriO}.

\subsection{A possible new cosmological model: $f(R)=R-{{ 6 \alpha} \over \sinh (R)}$  }

\noindent We consider moreover an alternative Lagrangian containing a term proportional to the inverse of the hyperbolic sinus of the generalized Ricci scalar:
\begin{equation}
f(R)=R-\frac{6 \alpha}{\sinh ( R)}
\end{equation}
where coefficients have been chosen to simplify  future equations and reproduce General Relativity in the appropriate limit.
Structural equations (\ref{struct}) hold in the form:
\begin{equation}
R-6 \alpha \frac{R \cosh(R)+2 \sinh(R)}{\sinh^2 (R)}=- \kappa \tau
\end{equation}
We consider the limit of the above structural equation for the case of small values of $R$ (see also \cite{Meng1} and \cite{Vollick}). Solutions can be found  under  the form:
\begin{equation}
R=\frac{- \kappa \tau \pm    \sqrt{(\kappa \tau)^2+72 \alpha (1+\alpha)}   }{2(1+\alpha)} \label{strshap}
\end{equation}
To reproduce, in the limit $\mid \kappa \tau \mid >> \alpha$ (see \cite{Vollick}), the result for General Relativity $R=- \frac{1}{1+\alpha} \kappa \tau$ we have to choose the plus sign in (\ref{strshap}). We remark that in the case $\alpha =0$ General Relativity is exactly recovered, as it should be expected. This choice implies that in the limit of late universe $\alpha >> \mid \kappa \tau \mid $ we will have:
\[
R \simeq   \sqrt{\frac{18 \alpha}{(1+\alpha)}}
\]
so that deviations from Einstein gravitational theory are large and the universe approaches in that limit a de-Sitter universe, dominated by a power-law expansion. \\
We obtain, in this case, that equation (\ref{MHC}) for the Hubble constant cannot be solved analytically; neither the structural equation (\ref{strshap}) provides an explicit analytical expression of R in terms of $\tau$. We can however write the system in an implicit form:
\begin{eqnarray}
H^2&=&\left[    \frac{-9(w+1) \alpha \kappa \sinh^2(R)(1+\cosh^2(R))}{(\sinh^2(R)+ 6 \alpha \cosh(R)) [ \sinh^3 (R)+ 6 \alpha (R+ R \cosh^2(R)+\sinh(R)\cosh(R))]}\right]^{-2} \times  \nonumber \\
& \times& \left[  \frac{\kappa \sinh^2(R)}{3[\sinh^2(R)+6 \alpha \cosh (R)]}   \left(  \rho+\frac{R-\frac{6 \alpha}{\sinh(R)}+\kappa \tau}{2 \kappa}  \right)\right] \label{53} \\
\nonumber\\
R&-&6 \alpha \frac{R \cosh(R)+2 \sinh(R)}{\sinh^2 (R)}=- \kappa \tau \label{54} 
\end{eqnarray}
where the solution $R=F(\tau)$ of (\ref{54}), substituted into (\ref{53}), gives a formal solution to the system. This expression could be possibly useful for some further numerical analysis.\\
Considering the leading terms in the limit of small $R$, corresponding to the present
universe we are interested in, the above system can be solved in a rough approximation. We obtain from the structural equation (\ref{strshap}) that $R=\frac{18 \alpha}{\kappa \tau}$ and substituting in the approximated expression for $H^2$ we obtain:
\begin{equation}
H^2\simeq [3w+4]^2 \left( -\frac{3 \alpha}{\kappa \tau} \right) \qquad \alpha>0 \label{prsinl}
\end{equation}
so that the current acceleration of the universe is explained by means of this model in correspondence with small values of $R$. As a matter of facts, apart from constant factors, we have that $H^2 \;  \simeq \;  \tau^{-1}$ and consequently for the most interesting case of dust universe we have that $H^2 \; \simeq \; a^3 (t)$, describing an accelerating universe.
In this cosmological model we will have that the effective $w_{eff}$ at present time, which means in the limit (\ref{prsinl}), can be simply obtained to be $w_{eff}=-2-w$, which goes in the direction of the experimental results of \cite{Spergel}.


\section{Conclusions and perspectives}
\noindent In this paper we have analyzed alternative theories of gravity, the Lagrangian of which is a general function of the generalized Ricci scalar $R$, constructed out of a dynamical metric $g$ and a dynamical connection $\Gamma$. The Palatini formalism provides first
order field equations for the metric and the connection $\Gamma$. A structural metric $h$ is introduced, such that the connection results to be the Levi-Civita connection of $h$ and $h$ is consequently conformal to $g$. Spacetime geometry is thus defined by means of generalized Einstein equations and it is in fact mutuated from structural equations which reproduce the standard Einstein spacetime with suitable choices of the parameters. \\
To treat explicitly cosmological models we choose $g$ to be a Robertson-Walker metric. This allows to obtain modified Friedmann field equations and a modified Hubble constant related to the conformal transformation between $g$ and $h$. The metric $h$ results to be FRW, too, so that it can be conveniently considered as
a physical metric in place of the original $g$.\\
If we moreover specialize to the pure-power  case  $f(R)=\beta R^n$ (with $n$ an arbitrary real exponent) we have seen that, with suitable choices of the parameters involved, these models are able to explain the current acceleration of the universe. We obtain  that polynomial Lagrangians in the generalized Ricci scalar provide an explanation for both present acceleration and inflation of the universe in  suitable limits \cite{Nojiri}.\\
Considering instead the case of logarithmic Lagrangians in $R$ we have that in the limit of small matter density, corresponding to late time universe, the solution approaches a de-Sitter universe with exponential power law expansion. This results into a possible model for accelerating universe. \\
We also propose a new possible cosmological model inlolving the inverse power of the hyperbolic sinus. Even if it is not  analytically solvable it however provides, in  a suitable limit, an explanation for the current acceleration of the universe as it should be expected.\\
The formalism developed here is very general and mathematically well defined. It provides useful physical results in the specific cases we have considered. Of course, a fully covariant satisfactory Lagrangian should be a more complicated polynomial-like expression involving more than one power  $R^m$ and more than one inverse power $R^{-n}$, besides a (possible) logarithmic factor $ln (\beta R)$ and terms proportional to the inverse power of $\sinh ( R)$. In this case, however, the non-linearity of the structure equation does not allow in general a simple analytical resolution as we did in the specific examples, so that numerical or approximation techniques have to be invoked for generic Lagrangians.\\
\\
Moreover, we stress that further applications of the formalism  developed here  can be extended to cover also generalized Ricci-squared theories, which  will be analyzed in a forthcoming paper  \cite{Ricci2}.

\section{Acknowledgements}
\noindent We are very grateful to Prof. S.D. Odintsov for useful discussions and very important suggestions, concerning the physical properties of the cases considered, and for suggesting us some of the examples.  \\
This work is partially supported  by GNFM--INdAM research project ``\emph{Metodi geometrici
in meccanica classica, teoria dei campi e termodinamica}'' and by MIUR: PRIN 2003 on
``\emph{Conservation laws and thermodynamics in continuum mechanics and field theories}''.
Gianluca Allemandi is supported by the I.N.d.A.M. grant: "Assegno di collaborazione ad attivit\'a di ricerca a.a. 2002-2003".

\end{document}